\documentclass[12pt,titlepage]{article}
\usepackage[letterpaper,top=1in,bottom=1in,left=1.25in,right=1.25in]{geometry}
\usepackage{graphicx,cite, url}
\usepackage{color}
\usepackage{hyperref}
\hypersetup{
    colorlinks=true,
    linkcolor=blue,
    filecolor=magenta,      
    urlcolor=blue,
}

\usepackage{amsmath,mathrsfs,cleveref,amsfonts}

\newcommand{\bw}{\mathbf{w}}
\newcommand{\bX}{\mathbf{X}}

\newcommand{\bY}{\mathbf{Y}}
\newcommand{\bx}{\mathbf{x}}
\newcommand{\by}{\mathbf{y}}

\newcommand{\cL}{\mathcal{L}}
\newcommand{\bbI}{\mathbb{I}}
\DeclareMathOperator{\var}{Var}
\DeclareMathOperator{\E}{\mathbb{E}}
\DeclareMathOperator{\acc}{Acc}
\DeclareMathOperator{\cred}{Cred}

\begin{document}

\title{Reliable deep-learning-based phase imaging with uncertainty quantification}
 
\author{Yujia Xue$^{1}$, Shiyi Cheng$^{1}$, Yunzhe Li$^{1}$, Lei Tian$^{1,*}$
\\
\multicolumn{1}{p{\textwidth}}{\centering\emph{\normalsize  Department of Electrical and Computer Engineering, Boston University, Boston, MA 02215, USA\\
		$^{*}$ leitian@bu.edu
}}}

\maketitle
\begin{abstract}
Emerging deep-learning (DL)-based techniques have significant potential to revolutionize biomedical imaging. 
However, one outstanding challenge is the lack of reliability assessment in the DL predictions, whose errors are commonly revealed only in hindsight.
Here, we propose a new Bayesian convolutional neural network (BNN)-based framework that overcomes this issue by quantifying the uncertainty of DL predictions.
Foremost, we show that BNN-predicted uncertainty maps provide surrogate estimates of the true error from the network model and measurement itself.
The uncertainty maps characterize imperfections often unknown in real-world applications, such as noise, model error, incomplete training data, and out-of-distribution testing data.
Quantifying this uncertainty provides a per-pixel estimate of the confidence level of the DL prediction as well as the quality of the model and dataset. 
We demonstrate this framework in the application of large space-bandwidth product phase imaging using a physics-guided coded illumination scheme.
From only five multiplexed illumination measurements, our BNN predicts gigapixel phase images in both static and dynamic biological samples with quantitative credibility assessment.
Furthermore, we show that low-certainty regions can identify spatially and temporally rare biological phenomena.
We believe our uncertainty learning framework is widely applicable to many DL-based biomedical imaging techniques for assessing the reliability of DL predictions.
\end{abstract}

\section{Introduction}

\begin{figure*}[t]
\centering
{\includegraphics[width=\linewidth]{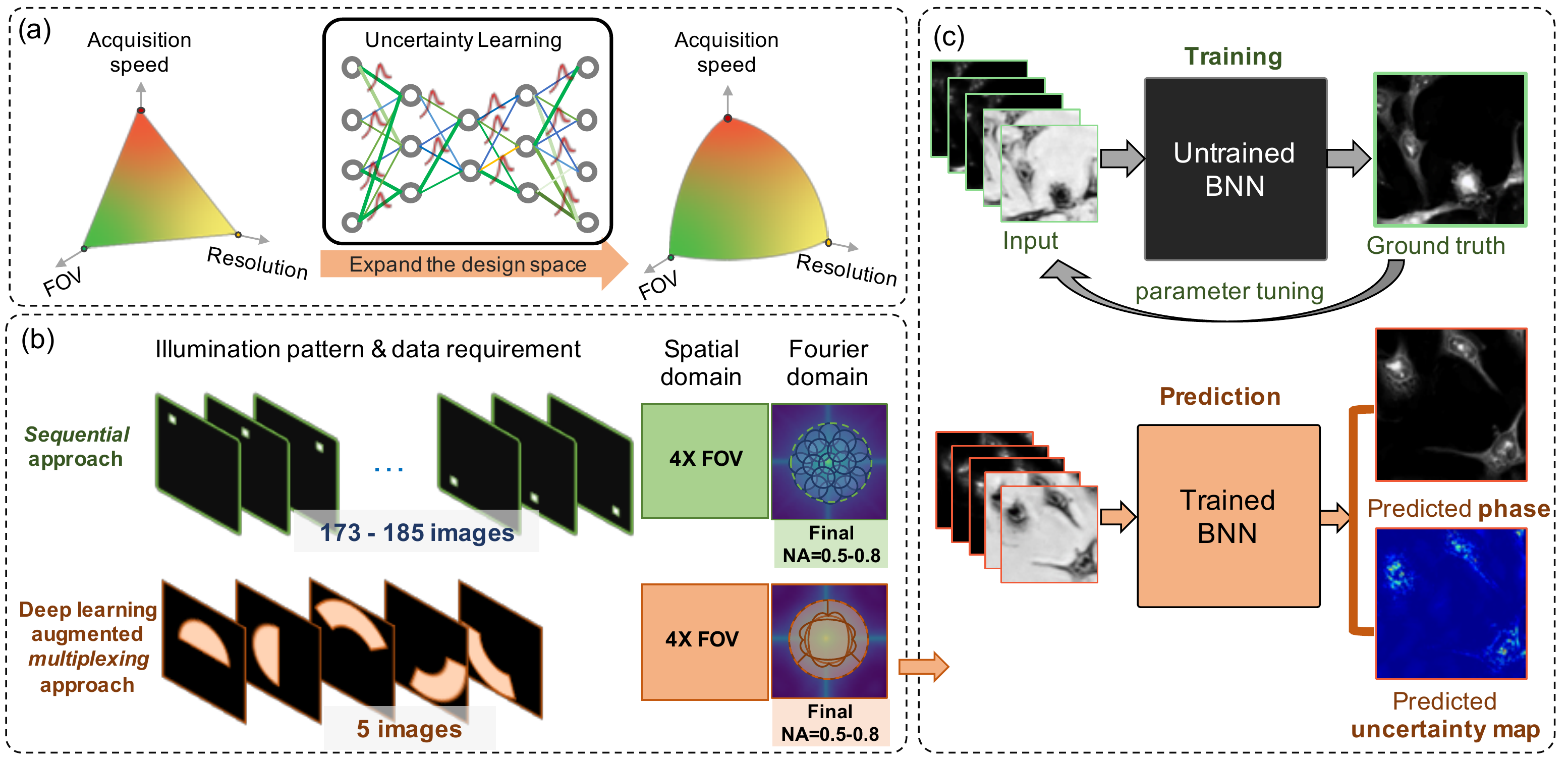}}
\caption{Overview of our reliable DL-based phase imaging technique. 
(a) Our technique opens up an expanded imaging attribute space, bypassing the conventional tradeoff between FOV, resolution, and acquisition speed. 
(b) It uses  five asymmetric  illumination coded intensities to encode large-SBP phase information.
(c) A BNN is developed to make phase predictions and quantify the uncertainties of the model. 
}
\label{fig:intro}
\end{figure*}

The imaging throughput of traditional techniques is fundamentally limited by the intrinsic trade-off among field-of-view (FOV), resolution, and acquisition speed.  
It is well known that the space-bandwidth product (SBP) of an optical system is invariant under any linear canonical transform~\cite{lohmann:96,lohmann1989}.  
Further considering super-resolution-type techniques that require multiple measurements, the acquisition time scales linearly with the expanded bandwidth in a single dimension, and quadratically for 2D isotropic resolution enhancement~\cite{Lukosz_1966,Lukosz_1967}. 
The same scaling law also applies to the scanning-based systems for enlarging the FOV.
Accordingly, the 3D trade-space spanned by the FOV, resolution, and acquisition speed can be visualized as shown in Fig.~\ref{fig:intro}(a), with a hyperplane defining the achievable imaging attributes which highlights the linear trade-off among them (for a 1D problem). 
The imaging techniques of our interest belong to the classical phase-retrieval problem. 
Despite the extra complexity from the intensity-only, nonlinear measurements, the general scaling law for the achievable imaging attributes follows the same trade-space, as studied both theoretically~\cite{wicker2014} and experimentally~\cite{zheng2013,tian2015b}.
Our first goal here is to investigate the feasibility of bypassing the classical limit imposed by the linear trade-space by combining non-conventional multiplexed measurement schemes and deep learning (DL). 
By doing so, our technique will open up an expanded design space that allows a combination of FOV, resolution, and acquisition speed beyond those achievable using conventional phase-retrieval techniques [as illustrated in Fig.~\ref{fig:intro}(a)]. 

Our work is inspired by the recent demonstration of several DL-based phase-retrieval techniques~\cite{sinha2017lensless,rivenson2018phase,Nguyen_2018,LiShuai_2018,Li_2018,Wu_2018,horstmeyer2017convolutional,Diederich_2018,Robey_2018,kellman2018physics}, which can be categorized into two classes.  
The first class focuses on solving the phase-retrieval problem alone using a convolutional neural network (CNN); no modification to the measurement procedure is made~\cite{sinha2017lensless,rivenson2018phase,Nguyen_2018,LiShuai_2018,Li_2018,Wu_2018}. 
As a result, these techniques generally do not improve the imaging throughput. 
Nevertheless, using the CNN-based algorithm has been reported to have several benefits, including its robustness to noise, scattering, and experimental errors~\cite{sinha2017lensless,rivenson2018phase,Nguyen_2018,LiShuai_2018,Li_2018,Wu_2018}. 
The second class focuses on introducing the physical model into the construction of the CNN.  
This is done by modeling the image formation process as the initial layers of the CNN~\cite{horstmeyer2017convolutional,Diederich_2018,Robey_2018,kellman2018physics}. 
As a result,  training  the CNN  jointly optimizes the physical parameters used in the acquisition alongside its computational parameters. 
However, the effectiveness of this approach relies on the accurate modeling of the image formation process~\cite{horstmeyer2017convolutional}, which can be difficult in practice due to the presence of uncalibrated aberrations and other experimental imperfections.

Differing from these two classes, we propose to solve the large-SBP phase-retrieval problem using a {\it physics-guided} DL approach, which consists of two complementary components. 
The first component is a highly measurement-efficient illumination multiplexing strategy designed by two physical principles.
First, we exploit asymmetric illumination to encode the phase information   into the intensity measurements based on the principle of differential phase contrast (DPC)~\cite{tian2015a}.
Second, we enhance the resolution following the principles of the synthetic aperture~\cite{hillman2009} and Fourier ptychographic microscopy (FPM)~\cite{zheng2013} by using oblique illumination to introduce into the measurements high-frequency information that are beyond the native passband of objective lens.
Most importantly, our method uses only {\it five} coded measurements regardless of the final resolution [Fig.~\ref{fig:intro}(b)], making our technique highly flexible and scalable for large-SBP phase-retrieval problems. 
As a result, our proposed technique avoids the need to quadratically increase the number of measurements to achieve a higher resolution; a limitation that is imposed by conventional FPM techniques. 
The reason behind preventing such multiplexed measurements to be used previously is the severe ill-posedness of the resulting inverse problem~\cite{tian2015b,tian2014,bostan2018accelerated,Chen_2018}.
This results in undesirable phase artifacts in the reconstruction from  existing multiplexed FPM (mFPM) algorithms.
The second component uses DL to overcome the ill-posedness of the inverse problem and complements the new measurement strategy.
Specifically, we show that our DL algorithm robustly inverts the physical model and recovers large-SBP phase information from highly  multiplexed nonlinear measurements, which would otherwise not be possible.

An important feature of our DL technique is the ability to quantitatively assess its reliability. 
In particular, we aim to address a common criticism on DL that the error of the prediction cannot be easily evaluated unless the ground truth is known. 
To address this issue, we develop an uncertainty learning (UL) framework based on the Bayesian convolutional neural network (BNN)~\cite{kendall2017uncertainties} [Fig.~\ref{fig:intro}(c)].
We show that the reliability of the BNN prediction can be quantified by two  predictive uncertainties, including the model uncertainty and the data uncertainty, akin to the epistemic  and aleatoric uncertainties, respectively, in Bayesian analysis~\cite{Kiureghian_2009}.
In particular, we show that the model uncertainty allows us to characterize the robustness of our physics-guided DL technique.
By training and testing on an ensemble of CNNs, the BNN quantifies the  variabilities intrinsic to the model without ``cherry-picking'' the results~\cite{kendall2017uncertainties}.
In addition, we show that the data uncertainty allows assessing the randomness of the predictions that originate from data imperfections~\cite{kendall2017uncertainties}, including noise, incompleteness in the training data, and the error due to out-of-distribution testing data.

In order to rigorously quantify the reliability of the BNN predictions, an important step is to perform statistical data analysis. 
We develop a procedure to relate the BNN output to  Bayesian statistical metrics, including credibility, credible interval, and reliability diagram.  
By doing so, our work establishes a comprehensive procedure for evaluating the reliability of our DL-based phase-retrieval technique.

By capturing experimental data on two different computational microscopy platforms,
we justify our proposition that our technique is applicable to different experimental setups. 
First, we demonstrate 5$\times$ resolution enhancement on the setup in~\cite{ling_2018}.
Next, we demonstrate the scalability of our technique by synthesizing multiplexed measurements on both static and dynamic biological data from~\cite{tian2015b} and achieve 4$\times$ resolution improvement. 
In addition, the robustness of our technique to common experimental factors is quantified by evaluating the BNN-predicted uncertainties, including spatially varying aberrations, illumination misalignment, and phase wrapping artifacts. 
Mostly importantly, the results show that the selection of the training data indeed affects the confidence of the prediction, whose effect can be quantified by our UL framework. 
Specifically, we investigate the effect of limited training data due to spatial and temporal constraints and biological sample types. 
Furthermore, the BNN is shown to be reliable when trained and tested on different sample types and under different experimental configurations.
The BNN-predicted uncertainties are shown to be indicative to the true error.
Finally, a  potential utility of our UL framework is explored in a time-series  experiment to identify rare biological structures and phenomena.

\section{Method}
\label{sec:method}
\subsection{Multiplexed illumination for large-SBP phase imaging}

Our illumination multiplexing scheme combines the physical principles of DPC~\cite{tian2015a} and FPM~\cite{zheng2013} to  encode high-resolution phase information across a wide FOV using a small number of intensity measurements. 
DPC is a phase microscopy technique that involves taking intensity measurements using asymmetric illumination~\cite{mehta2009}. 
Under the first Born approximation, a brightfield intensity measurement is linearly related to a sample's permittivity contrast by a weak phase transfer function~\cite{tian2015a}.  
The distribution of the transfer function affects the quality of the phase retrieval and can be tuned by adjusting the illumination pattern. 
Most importantly, the transfer function contains missing frequencies along the axis of  asymmetry for a given illumination pattern~\cite{tian2015a}.  
As a result, illumination patterns containing at least two axes of asymmetry are commonly used to ensure complete Fourier coverage. 
Several studies on the choice of illumination patterns have been performed based on the linear model~\cite{tian2015a,fan2019optimal}.
A CNN-based technique has also been developed to optimize the illumination patterns using a data-driven framework~\cite{kellman2018physics}.
It should be noted that the validity of the DPC model relies on the presence of a strong reference wave as in the brightfield measurements; the model no longer holds for darkfield measurements.
Accordingly, the maximum resolution achievable by DPC is limited to $2\times$ the objective NA. 

To further extend the resolution by more than $2\times$, our technique  adapts the principle of FPM. 
In FPM,  intensities are measured with  asymmetric illumination in both brightfield and darkfield.
Next, an iterative algorithm that simultaneously retrieves phase information and carries out the synthetic aperture is implemented.
As a result, this method can increase the resolution up to the sum of the illumination  and  objective NAs~\cite{zheng2013}.
A major advantage of FPM is its ability to achieve both a wide-FOV and a high resolution, i.e. a large SBP.  
However, its imaging throughput is limited by the long acquisition time imposed by the large data requirement.
Specifically, the original sequential FPM (sFPM) requires taking hundreds of images since it requires scanning through all the controllable illumination angles one by one~\cite{zheng2013} [Fig.~\ref{fig:intro}(b)]. 
The acquisition time can be shortened by illumination multiplexing in mFPM.
In~\cite{tian2014}, a random multiplexing scheme is shown to 
achieve up to 8$\times$ data reduction.
A hybrid multiplexing scheme that combines DPC in the brightfield with random multiplexing in the darkfield is shown to provide improved robustness in solving the mFPM phase-retrieval problem of mFPM~\cite{tian2015b}.
However, all these FPM schemes are fundamentally limited by the conventional trade-off, which results in an undesirable quadratic increase in the data requirement as the resolution increases~\cite{tian2015b}.

Here, we develop a DL-augmented  illumination multiplexing scheme that uses only five asymmetric illumination [Fig.~\ref{fig:intro}(b)].
First, we design two brightfield patterns based on the DPC model with in-total two axes of asymmetry (every $90^{\circ}$) to provide complete Fourier coverage within the brightfield limit. 
Next, we design three darkfield patterns with in-total three axes of asymmetry (every $120^{\circ}$) to further extend the Fourier coverage set by the sum of the illumination and  objective NAs, same as in the FPM. 
A notable feature of the proposed scheme is that extending the resolution simply requires modifying the illumination scheme to use a larger darkfield pattern, without the need for additional measurements. 
This means that the data requirement remains  the {\it same} as the resolution increases -- bypassing the limitation imposed by conventional techniques. 
By doing so, we improve the throughput of the data acquisition process by trading off computational complexity.
Specifically, the multiplexed measurements cannot be robustly inverted by existing model-based mFPM algorithms due to the severe ill-posedness of the inverse problem. 
We show that our proposed BNN-based algorithm overcomes this issue owing to its nonlinear multilayer structure.

\subsection{Uncertainty learning framework}

Our UL framework is built on the probabilistic view of  neural networks~\cite{Ghahramani_2015}.  
The learned neural network differs from training to training, which in turn results in varied predictions. 
The variability stems from several stochastic processes involved in the training, such as random weight initialization~\cite{pmlr-v9-glorot10a}, dropout~\cite{srivastava.etal2014}, and the stochastic-gradient-descent-type algorithms~\cite{Bottou_2010}. 
There are two ways to quantify the variabilities in a neural network, including the Bayesian~\cite{kendall2017uncertainties} and frequentist~\cite{lakshminarayanan2017simple} approaches. 
We outline both the approaches, provide the mathematical foundations for the Bayesian analysis, and then quantify uncertainties using both the Monte Carlo dropout~\cite{gal2016dropout} and the Deep Ensembles~\cite{lakshminarayanan2017simple}.  

\begin{figure}[t]
\centering
{\includegraphics[width=0.4\linewidth]{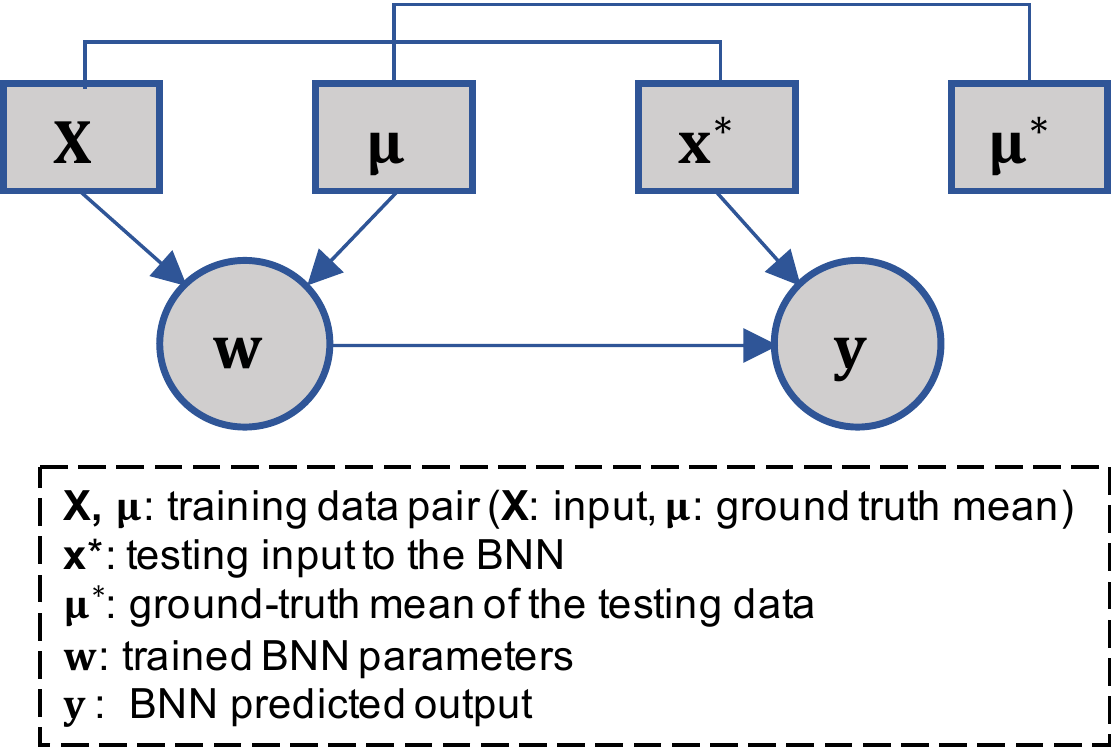}}
\caption{The graphical model of our UL framework that considers randomness in both the network weights $\bw$ and the predicted output $\by$.  
}
\label{fig:graph}
\end{figure}

\begin{figure*}[t]
\centering
{\includegraphics[width=0.95\linewidth]{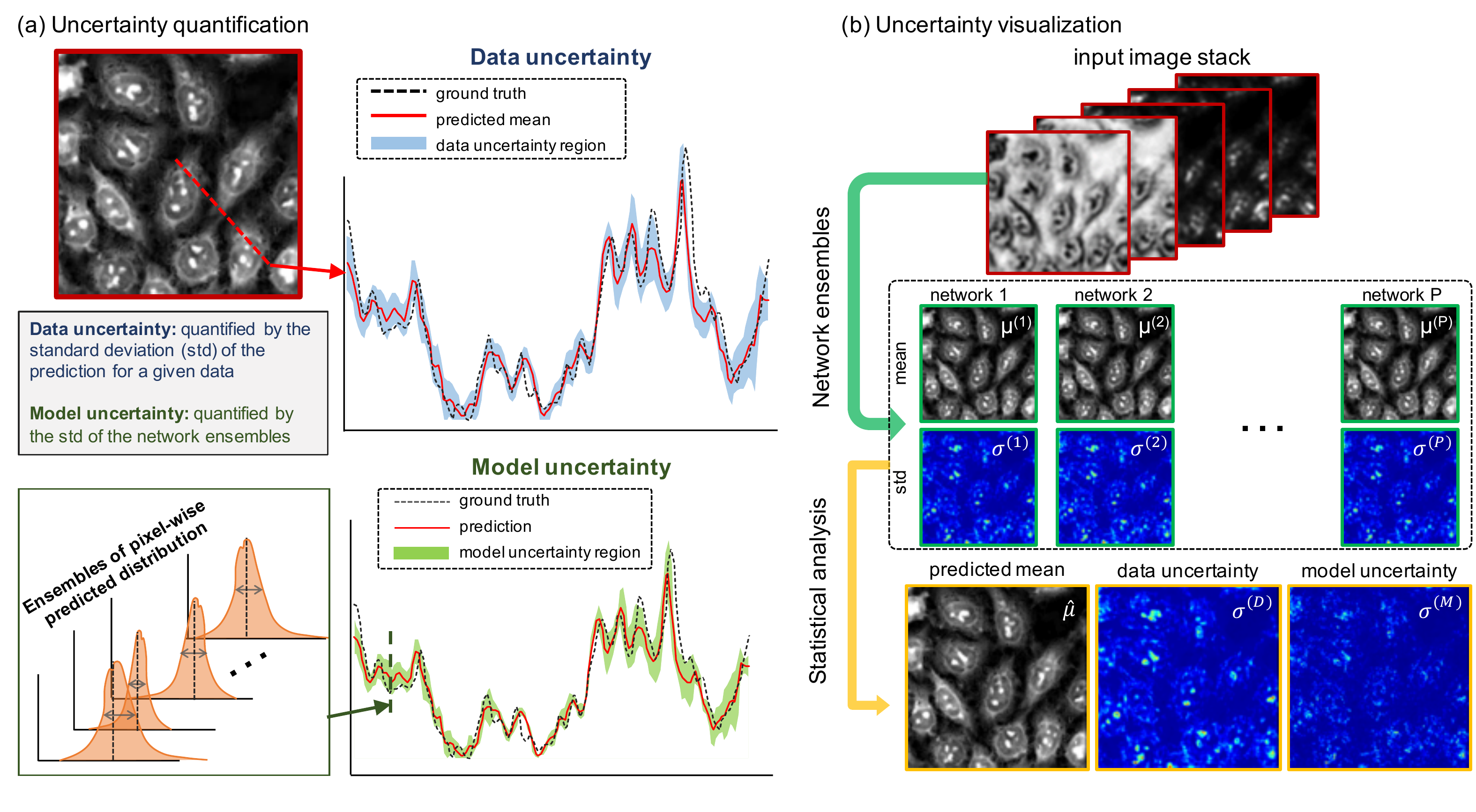}}
\caption{Overview of our UL framework. 
(a) The data uncertainty quantifies the effect of incomplete training data and is estimated via an uncertainty regularized loss function. 
The model uncertainty evaluates the stochasticity of neural network training and is estimated by network ensembles.
(b) During testing, the direct output from the BNN consists of an ensemble of mean and standard deviation maps.
Through statistical modeling, we obtain the final estimated phase, data and model uncertainty maps.}
\label{fig:uncertainty}
\end{figure*}

The BNN replaces the deterministic network weights with probability distributions over them [as illustrated in Fig.~\ref{fig:intro}(a)].
To quantify the variability of a prediction $\by$, we model the predictive distribution $p(\by|\bx^*, \bX, \bY)$ given the test input $\bx^*$ through marginalization over all the possible network weights~$\bw$ that were learned from the training data $(\bX, \bY)=\{\bx^t,\by^t\}^T_{t=1}$:
\begin{equation}
p(\by|\bx^*, \bX, \bY) =  \int p(\by|\bx^*,\bw)p(\bw|\bX, \bY)d\bw, 
\label{eq:p_predict}
\end{equation}
where Eq.~\eqref{eq:p_predict} applies the conditional independence between the training and testing data, and can be visualized by the graphical model in Fig.~\ref{fig:graph}. 
The posterior distribution $p(\bw|\bX, \bY)$ describes all the possible {\it network weights} given the training data.
The predictive distribution $p(\by|\bx^*,\bw)$ describes all the possible {\it predictions} given the network weights~$\bw$ and the testing input $\bx^*$ [Fig.~\ref{fig:uncertainty}(a) Top]. 
By modeling $p(\bw|\bX, \bY)$ and $p(\by|\bx^*,\bw)$, we can evaluate the model and data uncertainties, respectively.

To quantify the {\it data uncertainty}, we describe the probability distribution of the $k$th \mbox{$N$-pixel} random output of the BNN (given the input $\bx^k$) by a multivariate Laplacian distributed likelihood function:
\begin{eqnarray}
&& p(\by^k|\bx^k,\bw)=\prod_{i=1}^{N}{p(y^k_{i}|\bx^k,\bw)},\\
&& p(y^k_{i}|\bx^k,\bw) 
= \frac{1}{2\sigma^k_i}\exp\left(-\frac{|y^k_{i}-\mu^k_i|}{\sigma^k_i}\right),
\end{eqnarray}
where the output pixels (indexed by $i$) are assumed to be independent, and $\mu^k_i$ and $\sigma^k_i$ denote the pixel-wise mean and standard deviation, respectively.  
It can be shown that the widely used mean absolute error (MAE) corresponds to this Laplacian model with a constant standard deviation assumed for the entire output~\cite{kendall2017uncertainties}. 
By incorporating spatially {\it varying} standard deviations in our model, our BNN accounts for {\it inhomogeneous} noise and {\it shift-variant} model errors.

At the {\it training} stage, learning of the network weights is performed by minimizing  the normalized negative log-likelihood function, i.e. the loss function $L(\bw| \bx^t, \by^t)$, given the training data $(\bx^t, \by^t)$: 
\begin{equation}
L(\bw| \bx^t, \by^t) = \frac{1}{N}\sum_{i=1}^N 
\left[
\frac{|y^t_i-\mu^t_i|}{\sigma^t_i} + \log(2\sigma^t_i)
\right].
\end{equation}
$L(\bw| \bx^t, \by^t)$ consists of two parts:  
the first residual term resembles the MAE loss normalized by the pixel-wise standard deviation.
The second is the {\it data uncertainty regularization} term. 
Most importantly, one does {\it not} need the ground-truth mean ($\mu_i^t$) {\it nor} the ground-truth standard deviation ($\sigma_i^t$) for learning the uncertainty -- minimizing $L(\bw| \bx^t, \by^t)$ allows learning both using the sample pairs $(\bX, \bY)$ taken from the random process.
This is achieved by the structure of this loss function.
Specifically, a large residual error $|y^t_i-\mu^t_i|$ will be regulated by a large standard deviation, which in turn, increases the $\log(2\sigma^t_i)$ term; the optimum can only be reached when the two terms are balanced.
Training the BNN helps to not only find the optimal weights that explains all the data, but also quantifies the individual {\it mismatch} between the data and the model as measured by the spread ($\sigma^t_i$) in the network's output.
At the predication stage, the BNN estimates both the mean and the standard deviation given the testing input, as illustrated in Fig.~\ref{fig:intro}(c). 

One approach to assess the {\it model uncertainty} is to use the dropout network~\cite{gal2016dropout}. 
Briefly, with dropout applied before every weight layer,  a simple distribution $q(\bw)$ is learned to provide a variational Bayesian approximation to the posterior $p(\bw|\bX, \bY)$.

At the {\it prediction} stage, the model uncertainty is calculated by Monte Carlo dropout~\cite{gal2016dropout}.
By using Monte Carlo integration over $P$ samples satisfying $\bw^{(p)}\sim q(\bw)$, we can approximate the predictive distribution by a Laplacian mixture model:
\begin{eqnarray}
p(\by|\bx^*, \bX, \bY) \approx  \int p(\by|\bx^*,\bw)q(\bw)d\bw \approx \frac{1}{P}\sum_{p=1}^P p(\by|\bx^*,\bw^{(p)}).
\label{eq:MCDO}
\end{eqnarray}
The variations in the distributions $p(\by|\bx^*,\bw^{(p)})$ from the network ensembles are a consequence of the model uncertainty [Fig.~\ref{fig:uncertainty}(a) bottom]. 

The predicted mean $\hat{\mu}_i$ of the $i$th  pixel can be estimated by 
the unbiased minimum mean squared error estimator:
 \begin{equation}
\hat{\mu}_i \equiv \E [y_i|\bx^*, \bX, \bY]  \approx  \frac{1}{P}\sum_{p=1}^P\E [y_i|\bx^*,\bw^{(p)}] 
 \approx  \frac{1}{P}\sum_{p=1}^P 
 \mu_i^{(p)},
\label{eq:E}
\end{equation}
 where $\mu_i^{(p)}$  is the predicted mean from the $p$th network, and $\E$ denotes taking the expectation.

To provide a single, holistic measure of the uncertainty of the entire process, we quantify the overall uncertainty $\hat{\sigma}_i$ by computing the pixel-wise variance (Var):
\begin{equation}
\begin{split}
\hat{\sigma}^2_i \equiv &\var(y_i|\bx^*, \bX, \bY)\\
=& \E[\var(y_i|\bx^*, \bw)] + \var(\E[y_i|\bx^*, \bw])\\
  \approx &  \frac{1}{P}\sum_{p=1}^P 2\left(\sigma^{(p)}_i\right)^2
  + \frac{1}{P}\sum_{p=1}^P \left(\mu_i^{(p)}-\hat{\mu}_i\right)^2 \\
  = & \left(\sigma^{(D)}_i\right)^2 + \left(\sigma^{(M)}_i\right)^2 
\label{eq:var}
\end{split}
\end{equation}
where the first equality follows the law of total variance, and the second one is derived from Eq.~\eqref{eq:MCDO} and the Laplacian mixture model.
$\sigma^{(p)}_i$ denotes the    pixel-wise standard deviation predicted from the $p$th network ensemble.  
Equation~\eqref{eq:var} shows that the overall data uncertainty $\textstyle \sigma^{(D)}_i$ is measured by the mean of the  predicted variance; the model uncertainty $\textstyle \sigma^{(M)}_i$ is quantified by the variance of the predicted mean.

The second approach to quantify the uncertainties is the Deep Ensembles~\cite{lakshminarayanan2017simple}, in which multiple identical networks are trained under the same condition.
A sufficient number of trained networks fully captures the variabilities of the model. 
We train eight networks to quantify the uncertainties. 
The model uncertainty is quantified by the same procedures in Eqs.~(\ref{eq:E}, \ref{eq:var}).  

Some examples of the predicted mean phase map, data uncertainty map, and model uncertainty map are shown in Fig.~\ref{fig:uncertainty}(b). 
Comparisons between the Monte Carlo dropout and the Deep Ensembles are provided in the supplementary material.

\subsection{BNN structure}

\begin{figure}[t]
\centering
{\includegraphics[width=\linewidth]{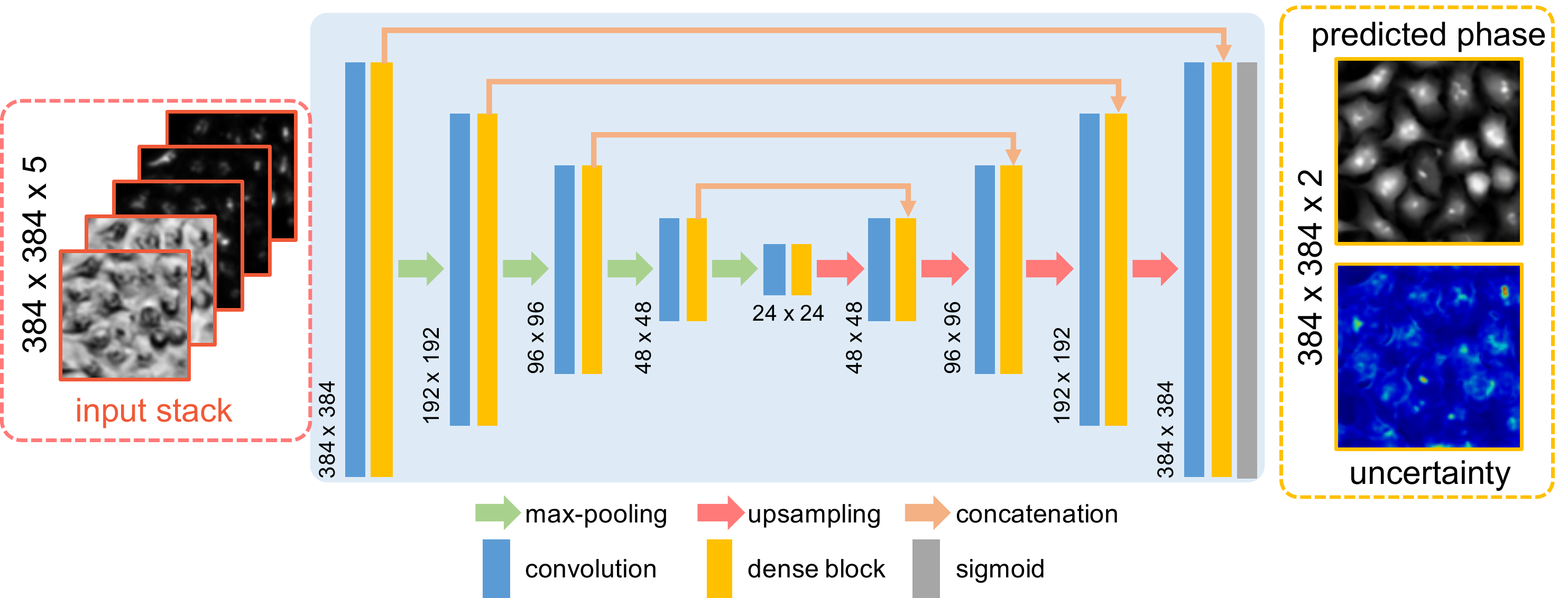}}
\caption{The BNN structure to perform UL.  
The main network  takes the U-Net structure.
The input takes the five low-resolution multiplexed intensity images.
The output predicts two-channel high-resolution phase and uncertainty maps.  
}
\label{fig:cnn}
\end{figure}

Our BNN follows the U-Net architecture owing to its versatility in solving image-to-image  problems~\cite{ronneberger2015u}.
It takes the encoder-decoder structure with skip connections to preserve high-frequency features, as shown in Fig.~\ref{fig:cnn}.
We made several modifications to perform uncertainty quantification.
Mostly importantly, the output of the BNN contains two channels, including the predicted (mean) phase map and the data uncertainty standard deviation map. 
To achieve high resolution enhancement, we further adapt the generative adversarial network (GAN)~\cite{isola2017image}.
We found that this GAN approach is needed to achieve 5$\times$ resolution improvement of data from our setup.
To achieve 4$\times$ resolution improvement, however, we do {\it not} need to use  the GAN . 
The impact of GAN on the reliability of the prediction is analyzed in Section~\ref{subsec:reliability}.  
Additional details about the network structure and training procedures are provided in the supplementary material. 
We have also made our implementation open source, along with pre-trained weights and test sample data, on our GitHub project page~\cite{git_UL_FPM}.

\subsection{Data acquisition}

Our technique is  tested on two LED-array-based computational microscope setups, detailed in~\cite{ling_2018,tian2015b} and five different types of biological samples. 
First, we collect data on unstained HeLa cells prepared under two fixation conditions, including ethanol and formalin, on the setup in~\cite{ling_2018}.
Depending on the fixation, unique morphologies can be observed in each sample, specifically in the plasma membrane and nuclei regions. 
All images are captured with a 4$\times$, 0.1 NA objective (Nikon CFI Plan Achromat).
Each data set  consists of the multiplexed data (two brightfield and three darkfield images) and the corresponding sFPM data (185 images).
Both the multiplexed and sFPM data are captured with the same 0.41 illumination NA, providing a final resolution of 0.51 NA .
Next, we validate our technique on the data from~\cite{tian2015b}.  
The multiplexed measurements are synthesized by summing the single-LED images.
We experimentally validate this procedure on the setup in ~\cite{ling_2018} and find that the numerically synthesized multiplexed intensity closely matches with the physically captured measurement since the LEDs are spatially and temporally incoherent.
We test our method on both fixed U2OS, MCF10A and dynamic live HeLa cell samples.
Images were captured with a 4$\times$, 0.2 NA objective (CFI Plan Apo Lambda)at an illumination NA of either 0.5 or 0.6, which provide a final NA of 0.7 -- 0.8.
Each dataset contains synthesized multiplexed and corresponding sFPM data.
More details are provided in the supplementary material.

\subsection{Training and test data configuration}
\label{sec:dataseparation}

We design three different training and testing data configurations in order to fully investigate the robustness of the BNN subject to different types of ``limited data'', including unseen biological sample types, a limited FOV, and inaccessible temporal data.

In the first set of experiments, the training data are taken from a single cell type; testing is then performed on several different cell types. 
In practice, different cells can produce out-of-distribution measurements that are not statistically ``similar'' to the training set.
Differing from the classification networks that are prone to testing errors from unseen object types, our network solves the inverse problem of an imaging model.  
As such, a properly trained network should be able to perform high-quality phase predictions and is robust to sample variations. 
We investigate how well the BNN can detect and quantify such abnormalities.  
In addition, we also study the network's robustness to variations in experimental setup.

In the second set of experiments, the training data are taken from a limited FOV region, whereas testing data are from the entire FOV. 
This task is of practical importance because wide-field systems like FPM often suffer from spatially variant aberrations~\cite{ou2014} and illumination mis-alignment~\cite{Eckert_2018}.
These variations in the imaging path can  change the intensity measurements significantly, such as contrast reversal, even when they are taken from the same sample, due to the interference effect. 
As a result, intensity measurements taken from FOV regions outside of the training region can produce out-of-distribution data due to the limited training FOV.
Differing from the model-based FPM approach, our data-driven BNN algorithm does {\it not} directly take any calibration information when constructing the network. 
Instead, the BNN needs to learn the spatially varying imaging model  from  the measurements and the ground truth phase.
We will investigate the reliability of the BNN against these model variations.

In the final set of experiments, the training data are taken from a limited observation time window from a time-series experiment. 
Dynamical biological processes can result in sample variations, which in turn affect the  statistics of the intensity measurements, which may be inconsistent with the training set.
 We will assess the BNN's ability to make  temporal predictions and quantify the uncertainty induced by the limited temporal data.

\subsection{Data preprocessing}
\label{sec:preprocessing}

To obtain the ground truth phase for training, we first perform phase reconstruction using the sFPM algorithm~\cite{tian2014}. 
To minimize model-mismatch-induced errors, we further perform algorithmic angle calibration using the algorithm in~\cite{Eckert_2018}, and digitally correct for the aberrations using the algorithm in~\cite{tian2014}.
Additional preprocessing is performed to remove the residual phase artifacts, including phase wrapping, a slowly varying background, and a large dynamic range. 
First, we perform phase unwrapping using the algorithm  in~\cite{Ghiglia_1994}.
Examples from this procedure are given in the supplementary material.
Next, the  slowly varying background artifact is removed with a morphological-opening-based algorithm.
Third, we perform phase dynamic range correction, which clips the $0.1\%$ pixels having extreme values to be a constant.
Finally, the phase is linearly normalized to $[0,1]$. 
This processed phase map is then cropped into small patches for training.
Still, the unwrapped phase contains residual isolated errors typically around large-phase or complex cellular features.
This results in incorrect ``phase labels'' in the training data, which later affects the prediction.
The impact of incorrect labels and phase clipping on the uncertainties of the phase predictions are analyzed in details in Sec.~\ref{subsec:scalability}.

To facilitate a later credibility analysis of the BNN output, we further quantify the noise present in the ground truth phase. 
Following~\cite{tian2015b}, we measure the standard deviation in the background region  and treat it as the intrinsic phase noise.
We assume that the same noise level is uniformly distributed also across the sample (e.g. cell) regions.
This noise level sets the tightest credible interval our BNN can provide; a detailed analysis is presented in Section~\ref{subsec:reliability}.

To preprocess the intensity measurements, background removal based on~\cite{tian2014,yeh2015} is first performed, followed by the dynamic range correction as in the ground truth phase preprocessing.
Next, the full FOV is divided into small patches, which are resized with a cubic interpolation algorithm to match the size of the input image with the ground truth phase. 
For training, the matching phase and intensity patches are fed into the BNN.
For testing, we apply an additional mean equalization to intensity patches taken from the {\it untrained} FOV region to alleviate the out-of-distribution effect.
We find this procedure is essential to improve the BNN's generalization.
Additional details about the preprocessing are provided in the supplementary material.

\subsection{Data analysis}
\label{sec:resultquant}

We develop data analysis procedures to quantitatively relate the BNN predictions to Bayesian statistical reliability measures.  
Typical neural networks  can only  evaluate errors based on  the ground truth, which is {\it not} possible for many practical problems.
Here, we derive a set of metrics that do {\it not} require knowing the ground truth.
Our analysis is based on the predictive Laplacian mixture model [Eq.~\eqref{eq:MCDO}]. The probability density of the $i$th pixel to take the value $y$ is 
\begin{equation}
 f_i(y)  \equiv p({y_i}=y|\bx^*,\bX,\bY) 
\approx \frac{1}{P}\sum_{p=1}^P \cL(y; \mu_i^{p}, \sigma_i^{p}).
\end{equation}
Accordingly, we define the {\it credible interval} $A^{\epsilon}_i  = [\mu_i-\epsilon,\mu_i+\epsilon]$ and its bound $\epsilon$.
The corresponding {\it credibility} $p^{\epsilon}_i$ is the predicted probability that the true mean $\mu^*_i$ falls within $A^{\epsilon}_i$: 
\begin{equation}
\begin{split}
p^{\epsilon}_i  & \equiv  g_i(\epsilon)
=\int_{\mu_i-\epsilon}^{\mu_i+\epsilon}f_i(y)dy 
 \\
& =  \frac{1}{P}\sum_{p=1}^{P}\left[
F^{p}(\mu_i+\epsilon)-F^{p}(\mu_i-\epsilon)
\right], \mathrm{for}~y_i\in A^{\epsilon}_i
\end{split}
\label{eq:conf}
\end{equation}
where $F^{p}(\cdot)$ is the cumulative distribution function (CDF) of the $p$th predicted Laplace distribution from the neural network ensembles. 
Another way to quantify the reliability is to calculate the bound $\epsilon^p_i$ given a targeted credibility $p$ and the predictive Laplacian mixture model; this can be computed by using the inverse function $g_i^{-1}(\cdot)$:
\begin{equation}
\epsilon^p_i=g_i^{-1}(p).
\label{eq:error}
\end{equation}
$g^{-1}(\cdot)$ does not have an elementary function, which is approximated by the bisection method.

To ensure the predictive metrics in Eqs.~(\ref{eq:conf}, \ref{eq:error})
are indicative, we further characterize how well they are {\it calibrated}~\cite{kuleshov2018accurate}.
To quantify this, a standard procedure is to compute the {\it reliability diagram} that compares the {\it accuracy}, i.e. the empirical probability of the ground truth matching with the predicted value, and the {\it credibility}~\cite{Niculescu_Mizil_2005}. 
Well-calibrated metrics should predict credibility similar to the accuracy -- the reliability diagram is diagonal. 
For the regression problem like ours, we adapt the modified reliability diagram~\cite{Weigert_2018}, which compares the averaged credibility and the empirical accuracy.
To generate a reliability diagram with $M$ probability bins, we define the bin interval $\Delta p = 1/M$ and the $m$th bin $P_m$ bounded by $p_{m-1}$ and $p_m$.
The averaged credibility $\cred({P_m},{\epsilon})$, takes the mean over the set of pixels $S^{\epsilon}_m$ having similar credibility within $(p_{m-1},p_m]$:
\begin{equation}
\cred({P_m},{\epsilon}) = 
\frac{1}{|S^{\epsilon}_m|}
\sum_{i\in S^{\epsilon}_m} p_i 
= 
\frac{1}{|S^{\epsilon}_m|}\sum_{p_i\in (p_{m-1},p_m]} p_i, 
\label{eq:confmetric}
\end{equation}
where $|S^{\epsilon}_m|$ measures the total number of pixels within the set.  
The empirical accuracy $\acc({P_m},{\epsilon})$ is defined as the fraction (empirical probability) of the pixels in set $S^{\epsilon}_m$ in which the ground truth mean~$\mu^*_i$ is within the corresponding credible intervals $A^{\epsilon}_i$:
\begin{eqnarray}
\acc({P_m},{\epsilon})
= \frac{1}{|S^{\epsilon}_m|}
\sum_{i\in S^{\epsilon}_m}
\bbI_{\{\mu^*_i\in A^{\epsilon}_i\}} 
= \frac{1}{|S^{\epsilon}_m|}
\sum_{i\in S^{\epsilon}_m}
\bbI_{\{\mu_i-\epsilon\leq \mu^*_i\leq \mu_i+\epsilon\}}.
\label{eq:acc}
\end{eqnarray}
In practice, the ground truth mean is unknown and can only be approximated by the sFPM phase that is ``noisy'', and so $\acc({P_m},{\epsilon})$ is influenced by the quality of sFPM reconstruction.
The bin interval sets the sampling interval in the reliability diagram, and also affects the sample size in $S^{\epsilon}_m$.  
We use the minimum interval while ensuring sufficient sample size for reliable statistical calculation. 
Both the averaged credibility and empirical accuracy depend on the credible interval bound $\epsilon$.  
We assess our model using different $\epsilon$ values in Section~\ref{subsec:reliability}.

\section{Results}
\label{sec:results}

Our results are presented in the following order: 
first, we show that our technique provides high-resolution phase predictions, and that the  uncertainty maps are highly indicative to the true error. 
In addition, we show that the method is scalable to different sample types, and  is applicable to experimental setups with varying final resolution. 
Second, we present large-SBP phase prediction and show that the uncertainty maps allow quantifying the effect of out-of-distribution data due to a limited FOV.
Third, we establish the reliability of our technique by performing statistical analysis.
Finally, we demonstrate time-series predictions and show that UL can facilitate the discovery of spatially and temporally rare biological features and events.

\subsection{Scalable illumination coding based DL phase imaging}
\label{subsec:scalability}

\begin{figure*}[t]
\centering
{\includegraphics[width=\linewidth]{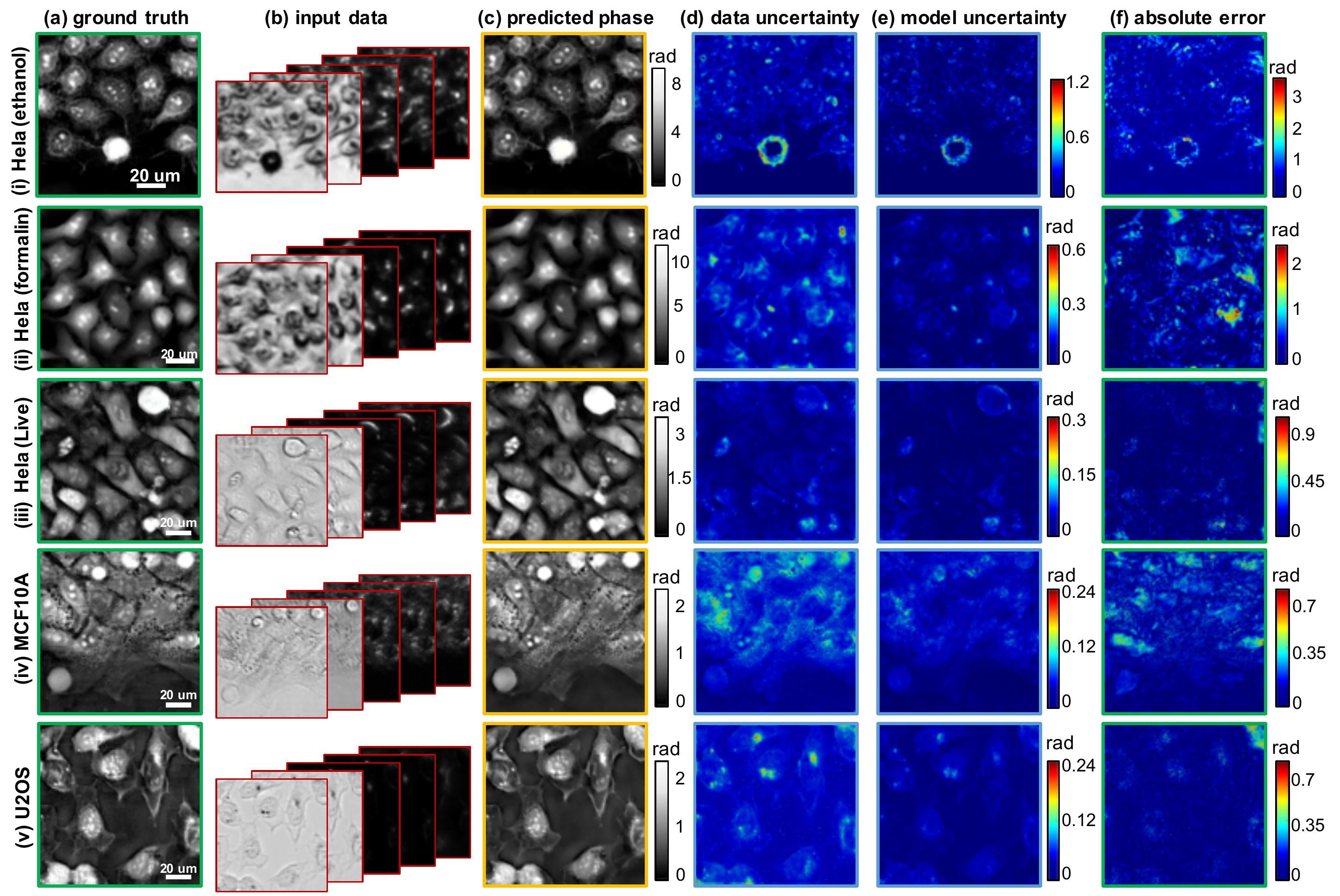}}
\caption{High-resolution phase estimation from DL-augmented coded measurements.
(a) The ground truth phase obtained from the sFPM.
(b) The input to the neural network consists of five low-resolution intensity images, including two brightfield, three darkfield.
Our BNN prediction includes (c) phase, (d) data uncertainty, and (e) model uncertainty. 
(f) The absolute error is calculated between the predicted and the ground truth phase. 
The uncertainty maps are highly correlated with the error maps, demonstrating the predictive power of our UL framework.
Unlike existing FPM techniques, our method requires the same number of measurements when the final resolution increases. 
Hela cells fixed in (i) ethanol and (ii) formalin are imaged with a 4$\times$ 0.1 NA objective and reconstructed with 0.5 NA resolution.
(iii) Live Hela and (iv) fixed MCF10A cells are imaged with a 4$\times$ 0.2 NA objective and reconstructed with 0.8 NA resolution.
(v) fixed U2OS cells are imaged with a 4$\times$ 0.2 NA objective and reconstructed with 0.7 NA resolution.
}
\label{fig:phase}
\end{figure*}

Our illumination coding scheme is highly scalable to large-SBP applications since 
it always uses five multiplexed measurements for achieving different resolution.
Experiments are performed on five cell types capture with two microscope setups and three different resolutions are achieved.
Specifically, Fig.~\ref{fig:phase}(i) and (ii) are obtained with the setup in~\cite{ling_2018} and achieve resolution enhancement from 0.1 NA to 0.51 NA; Fig.~\ref{fig:phase}(iii--v) are from the setup in~\cite{tian2015b}; (iii) and (iv) enhances resolution from 0.2 NA to 0.8 NA, and (v) from 0.2 NA to 0.7 NA.
First, we present results from training individual network for each cell type.
Without any hyper-parameter tuning, the same network structure is applicable to different samples captured on different setups.
Next, we show that the  BNN trained with a single cell type is generalizable to other ``unseen'' cell types.

Example multiplexed intensity measurements are shown in Fig.~\ref{fig:phase}(b). 
Our BNN is able to consistently provide high-quality phase predictions, as shown in Fig.~\ref{fig:phase}(c).
To evaluate a BNN predicted phase, we first compare it with the phase from sFPM in Fig.~\ref{fig:phase}(a) and then compute the pixel-wise absolute error map in Fig.~\ref{fig:phase}(f). 
Adding the additional uncertainty prediction in the BNN does {\it not} degrade the phase predictions as compared with the CNN approach (see supplementary material).
To demonstrate the need for using the DL method to overcome the ill-posedness of the phase-retrieval problem, we compare our results with those from two state-of-the-art model-based algorithms using the same multiplexed measurements.
The linear DPC model~\cite{tian2015a} can only recover phase with limited resolution, whereas the mFPM algorithm~\cite{tian2015b} results in high-frequency artifacts in the recovered phase (see supplementary material).

\begin{figure*}[t]
\centering
{\includegraphics[width=0.75\linewidth]{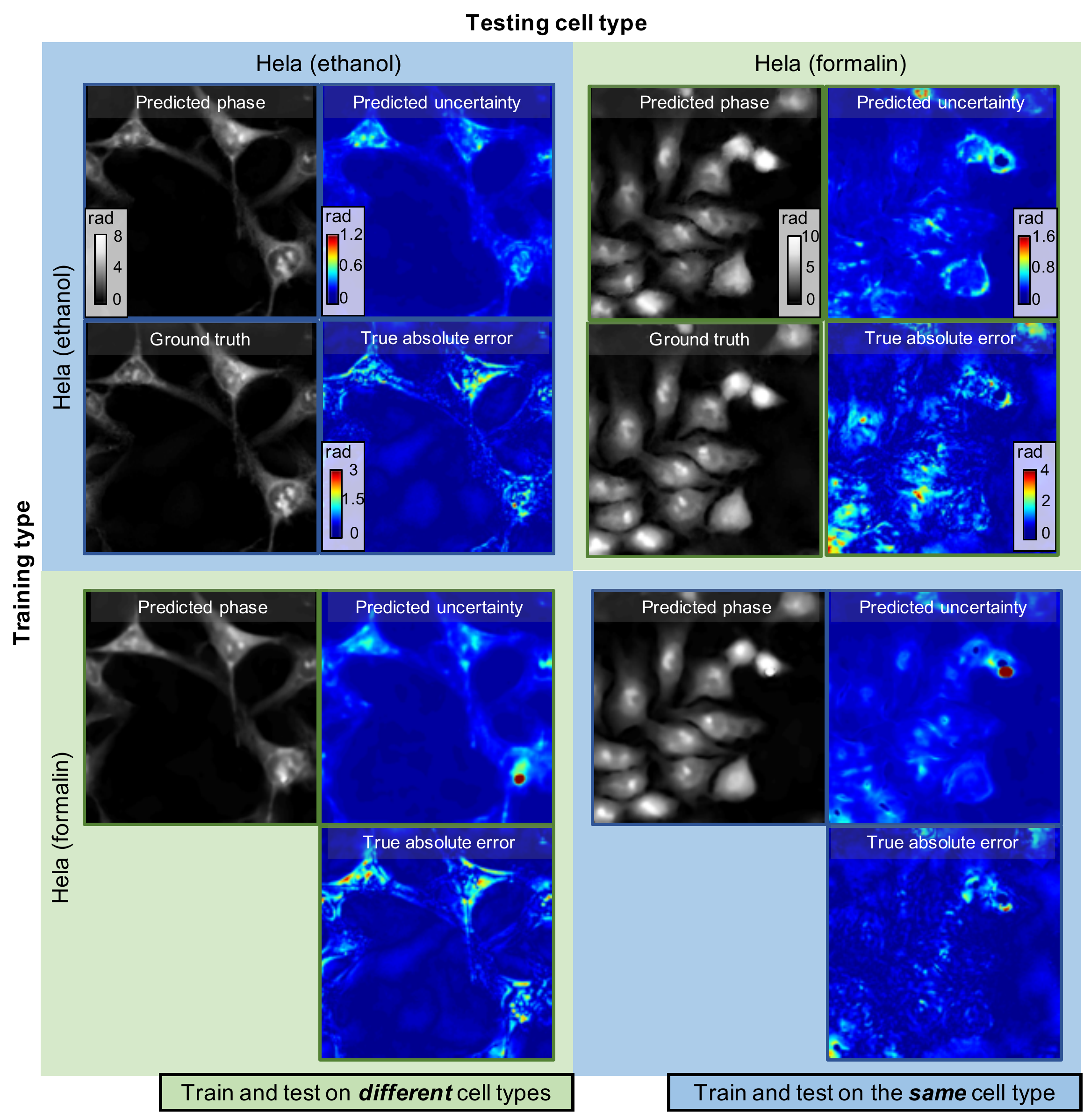}}
\caption{BNN predictions under different training and testing data configurations. 
The network can robustly perform phase retrieval against variations in the sample type. 
The uncertainty map can reliably detect potential errors in the phase predictions and is consistent with the true error.
}
\label{fig:crossval}
\end{figure*}

Next, we inspect the BNN-predicted data uncertainty [Fig.~\ref{fig:phase}(d)] and model uncertainty maps [Fig.~\ref{fig:phase}(e)].
The regions where the BNN {\it potentially} makes larger errors are marked with higher uncertainties. 
We observe that the uncertainty maps generally match well with the corresponding absolute error map.
In addition, the  predicted uncertainty values are about $1/3$ of the absolute error. 
This is because for a Laplace distribution ``$3\sigma$'' closely approximates the credible interval bound with $95\%$ credibility.
This demonstrates the utility of the uncertainty maps as a direct measure of the accuracy of the neural network predictions.
 Further quantitative reliability analyses are discussed in Section~\ref{subsec:reliability}.
In addition, we observe that the data uncertainty is the dominant term in our experiments, which suggests that the incompleteness in the training data is the main source of  error in the prediction.
Indeed, our training data are only taken from a small region of the FOV, as further discussed in Section~\ref{subsec:fullFOV}.
The low model uncertainty indicates that the predicted phase (i.e. pixel-wise mean) does not vary much across different neural network ensembles. 
This suggests that phase predictions based on the multiplexed measurements can be performed {\it consistently} -- the stochastic training process does {\it not} lead to unstable  inference results.
Furthermore, the high uncertainty regions consistently correspond to the cellular features with large phase values.
We attribute this to two primary sources of error.
First, the phase clipping inevitably introduces unwanted  saturation artifacts in the ground truth phase.
Second, although we correct for phase wrapping artifacts when generating the ground truth, residual errors still exist. 
Due to the presence of these inconsistencies present in the training data associated with the large-phase features, the trained BNN tends to flag such ``abnormal'' regions in the uncertainty output.

Our BNN is trained to solve an inverse problem. 
As such, a properly trained network learns to invert the physical model, which is independent of the type of objects used in the training. 
To justify this proposition, we compare the results from the BNN trained from the same cell type and from a different cell type in Fig.~\ref{fig:crossval}.
In general, the BNN is  able to make high-quality phase predictions and is robust to the selection of the sample type. 
Nevertheless, a slight degradation is observed in the phase predicted from the network trained from a different cell type.
This is because different cell types have distinct morphological features that can result in different intensity measurements.  
If the training data do not fully capture the statistical variations in the measurements, less accurate phase predictions would be produced when the network input contains ``out-of-distribution'' measurements.
Most importantly, the uncertainty map from the BNN can automatically detect such abnormalities in the data.  
As highlighted in Fig.~\ref{fig:crossval}, the uncertainty map remains highly indicative to the true absolute error regardless of the cell types being used for training and testing.
Additional results to demonstrate the robustness of our BNN to both sample  and setup variations are provided in the supplementary material.

\subsection{Large-SBP phase prediction and uncertainty quantification}
\label{subsec:fullFOV}

\begin{figure*}[t]
\centering
{\includegraphics[width=\linewidth]{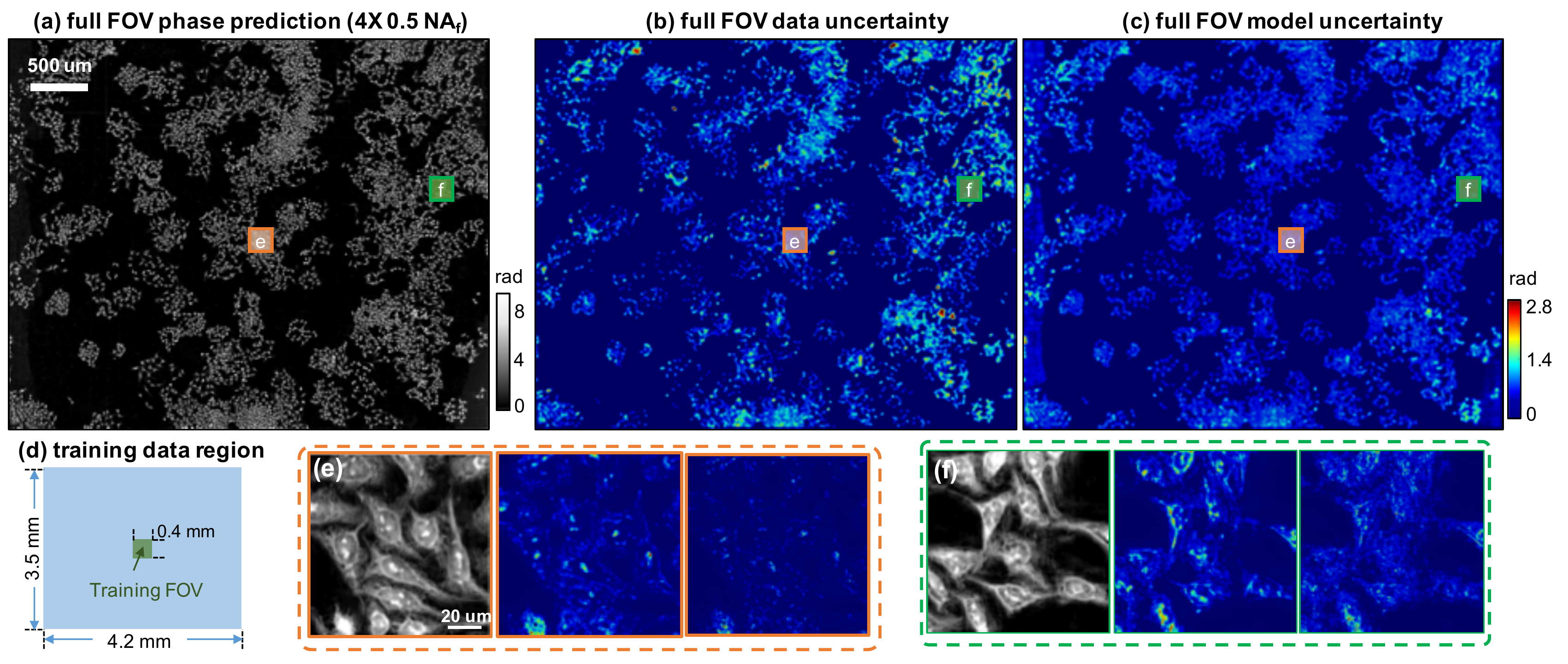}}
\caption{Large-SBP phase prediction and uncertainty quantification. 
(a) Full-FOV phase prediction achieving 0.51 NA resolution across a 4$\times$ FOV. 
(b) The data uncertainty map reliably identifies the out-of-distribution data corresponding to the peripheral FOV regions.
(c)~The model uncertainty is consistently low across the FOV except around the boundary, validating the robustness of our model.
(d)~The training data is  taken only from the central $0.4\times0.4$mm$^2$ region. 
Zoom-in of the predicted phase, data uncertainty, and model uncertainty of the region from (e) the central FOV and (f) the outer FOV. }
\label{fig:fullfov}
\end{figure*}

Next, we present large-SBP phase prediction across a wide FOV. 
Our BNN is trained on small image patches.
We perform phase and uncertainty predictions patch-by-patch. 
The full-FOV predictions in Fig.~\ref{fig:fullfov}(a--c) are obtained by stitching the patches  using the alpha blending algorithm. 

The full-FOV {\it model} uncertainty [Fig.~\ref{fig:fullfov}(c)] allows critically assessing the robustness of our technique. 
We observe that the model uncertainty is low across the FOV except in small regions around the boundary.
This verifies that the BNN can reliably make high-resolution phase predictions from the multiplexed measurements -- the predicted mean does not vary much across different network ensembles.
In the boundary regions, the measurements suffer from severe experimental errors that lead to  higher variations in the predicted means.


 
The effect of the out-of-distribution data due to the limited FOV is studied as follows:
our {\it training} data are taken from a small central region ($0.4\times0.4$mm$^2$ from the full $3.5\times4.2$mm$^2$ FOV), as shown in Fig.~\ref{fig:fullfov}(d).
In general, aberration degrades as the field angle increases (i.e. the distance away from the center).
In addition, the LED illumination produces greater angle mis-calibration~\cite{Sun_2016} and background non-uniformity as the field angle increases. 
Both effects imply a greater degree of out-of-distribution as compared with the training data. 
Importantly, {\it our UL approach allows predicting the potential errors induced by the out-of-distribution data} -- the data uncertainty map predicts higher standard deviation at the peripheral FOV regions [Fig.~\ref{fig:fullfov}(b)]. 

Identifying such data incompleteness {\it a posteriori} provides important feedback  to improve the data pipeline in DL. 
Intuitively, introducing previously out-of-distribution data to the training can reduce the data uncertainty. 
In our case, more credible predictions can be made by training on more examples encompassing aberrations and angle miscalibration in other FOV regions, as verified by additional experiments detailed in the supplementary material.

\subsection{Quantitative reliability analysis}
\label{subsec:reliability}

\begin{figure*}[t]
\centering
{\includegraphics[width=0.95\linewidth]{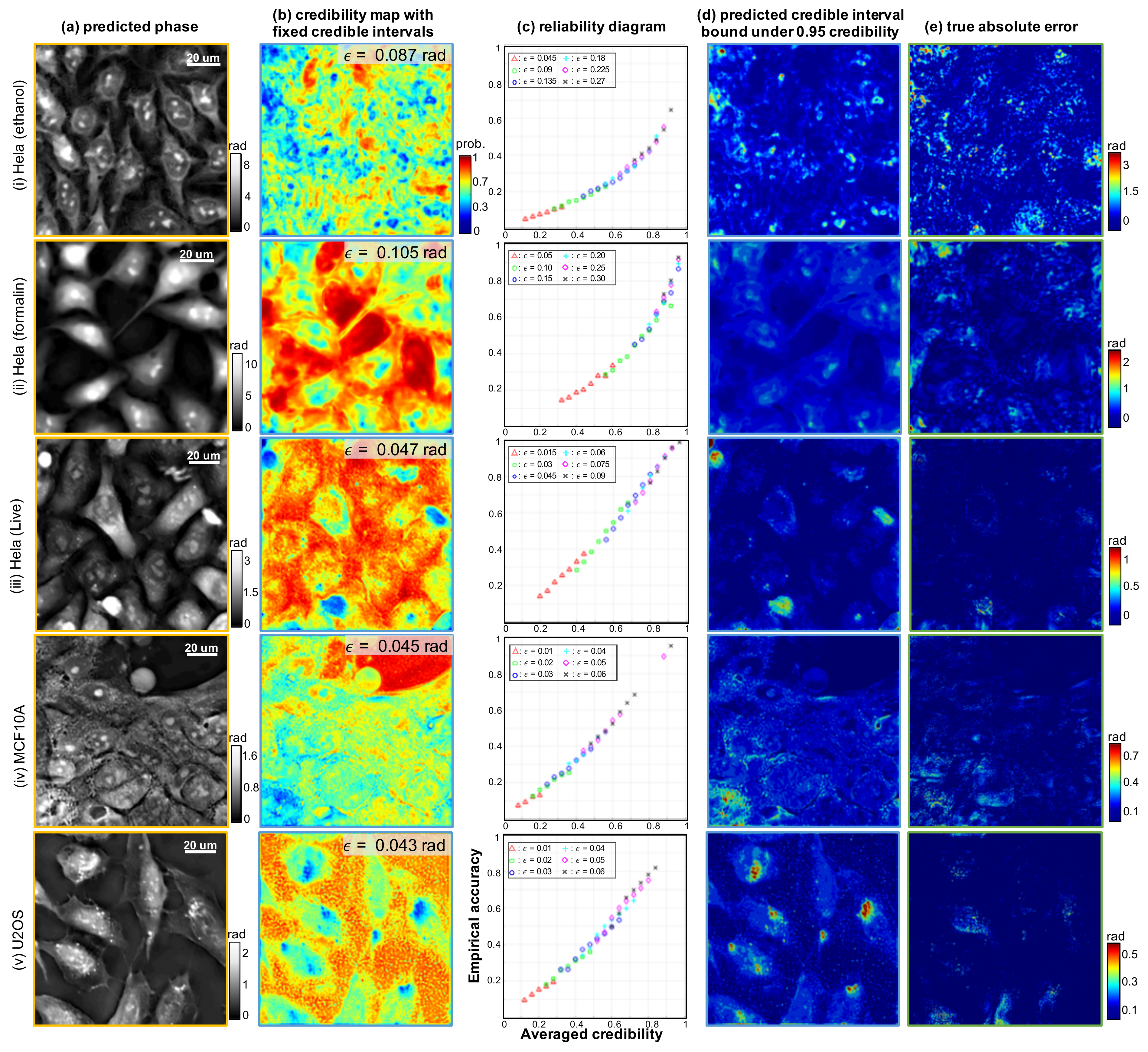}}
\caption{Reliability analysis of our predictions. 
(a) The predicted phase. 
(b) The credibility map calculated under the credible interval bound set by the intrinsic noise in sFPM. 
The less credible regions match with the out-of-distribution data containing phase clipping and wrapping artifacts. 
(c)  The reliability diagram computed by comparing the averaged credibility and empirical accuracy show that  (i--ii) are slightly over-confident, and (iii--v) are well calibrated. 
(d) The predicted credible interval bound under 95\% credibility correlate well with the corresponding true absolute error in (e). 
}
\label{fig:analysis}
\end{figure*}

To provide a quantitative assessment to our prediction, we first calculate the {\it credibility map} from the predicted pixel-wise distribution.
Given the bound $\epsilon$ and the predicted mean $\mu_i$ (at pixel $i$), the credibility $p_i^{\epsilon}$ [Eq.~\eqref{eq:conf}] measures the BNN-predicted probability that the true mean falls in the credible interval $A_i^{\epsilon}=[\mu_i-\epsilon,\mu_i+\epsilon]$. 
To properly choose $\epsilon$, we consider the intrinsic noise in the sFPM reconstructed phase by measuring the background standard deviation $\sigma_{\mathrm{background}}$.
We take this sFPM noise level as the credible interval bound ($\epsilon=\sigma_{\mathrm{background}}$) and compute the credibility pixel-by-pixel.
The credibility map provides a direct quantification of how much one can trust the BNN predicted phase. 
The credibility maps for the five samples and the credible interval bounds are shown in Fig.~\ref{fig:analysis}(b).
As expected, less credible regions point to the ``abnormal'' regions where phase clipping or wrapping artifacts are likely present in the training data.

Alternatively, we evaluate the credible interval bound giving a desired credibility.
The bound $\epsilon_i^p$ (at  pixel $i$) is computed using Eq.~\eqref{eq:error}.
By setting a constant $p=0.95$ (i.e. 95\%) credibility across the whole image, we compute the predicted credible interval bound map in Fig.~\ref{fig:analysis}(d).
We observe that the credible interval bound map generally encompasses the corresponding  true absolute error map [Fig.~\ref{fig:analysis}(e)].
These results match well with our previous observations on the predicted uncertainty maps.

Finally, we assess how well our UL framework is calibrated. 
We generate the reliability diagram [Fig.~\ref{fig:analysis}(c)] by computing the averaged credibility [Eq.~\eqref{eq:confmetric}] and the approximated accuracy [Eq.~\eqref{eq:acc}].
We set the probability bin interval $\Delta p = 0.04$ and use six credible interval bounds ($\epsilon$).
The first two cases [Fig.~\ref{fig:analysis}(i--ii)] {\it with} GAN included both show slightly over-confident predictions, as indicated by the curves below the diagonal.
The other three cases [Fig.~\ref{fig:analysis}(iii--v)]  {\it without} GAN provide better calibrated predictions since the curves closely follow the diagonal.
Besides the difference in the BNN structures, the first two cases have $\sim 3\times$ stronger phase resulting in more phase clipping induced errors, and $\sim 2\times$ higher intrinsic noise in the ground truth.
Since  the estimated empirical accuracy is also influenced by the quality of ground truth, the lower quality ground truth phase in the first two cases could also contribute to the less calibrated predictions. 
Methods to improve the calibration of BNN is an active area of research~\cite{kuleshov2018accurate} and will be developed in our future work.

\subsection{Time-series large-SBP phase and credibility prediction}
\label{subsec:time}

\begin{figure*}[t]
\centering
{\includegraphics[width=\linewidth]{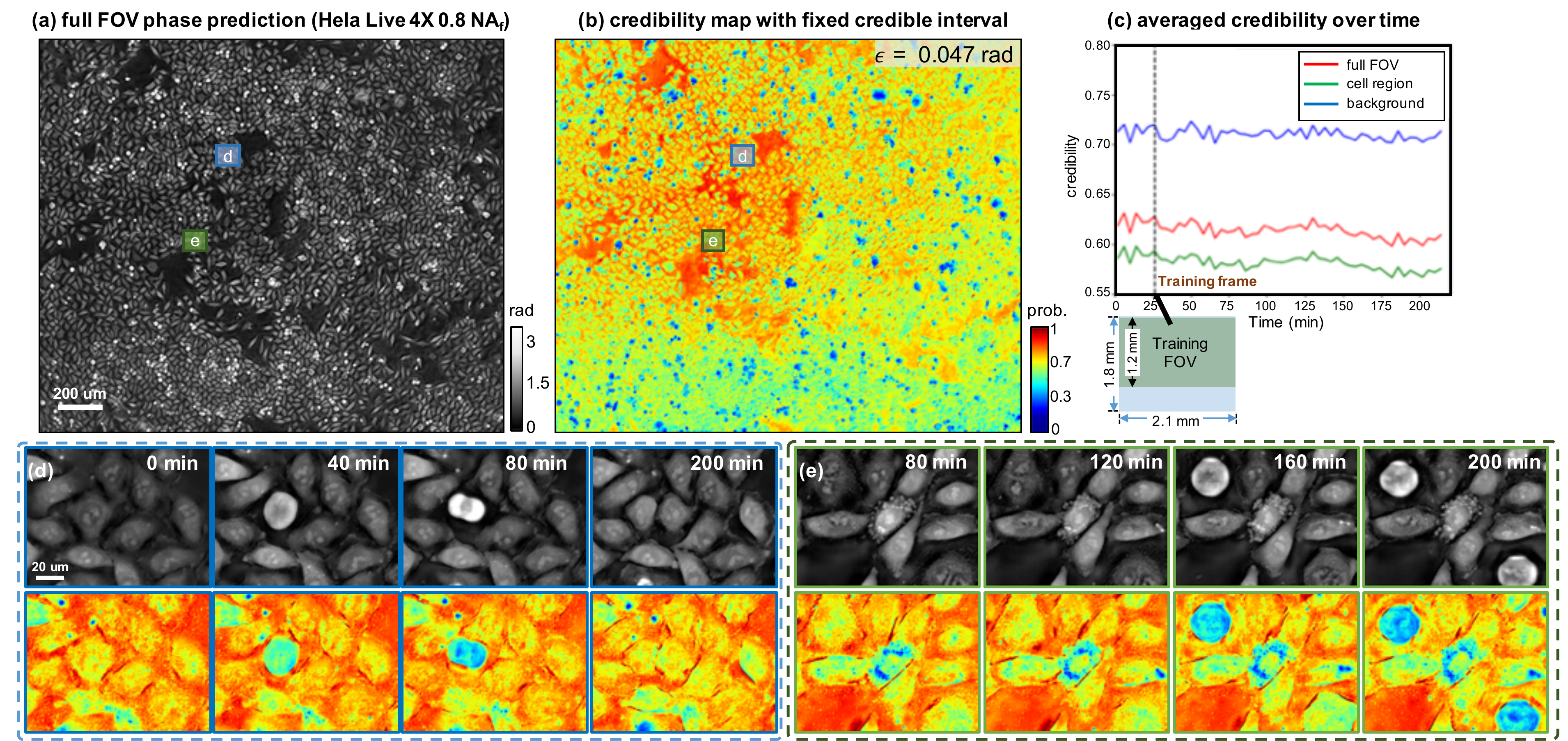}}
\caption{Time-series phase and credibility prediction. 
A representative frame from (a) the full-FOV  phase prediction achieving 0.8 NA across  4$\times$ FOV and (b) credibility map with a credible interval bound $\epsilon=0.047$~rad. 
(c) The training data is taken from the upper 3/4 of the FOV at the 26min frame. 
The averaged credibility are calculated over time on the whole FOV (red), the cell region (green), and the background (blue). 
This allows quantifying the ``temporal decorrelation'' induced by the temporal dynamics.
(d--e) Spatially and/or temporally rare events including cell mitosis and apoptosis, result in out-of-distribution data during prediction are automatically discovered by our BNN.  
The full time-series prediction is provided in the movie in Visualization 1.
}
\label{fig:conf_time}
\end{figure*}

Our technique is also applicable to imaging dynamic samples. 
Fig.~\ref{fig:conf_time} shows time-series predictions made by training the BNN using data only from a single time frame. 
We train the BNN  using the upper $3/4$ of the FOV at the 26min frame and perform full-FOV predictions on the rest of time frames.
An example FOV phase prediction is shown in Fig.~\ref{fig:conf_time}(a).
The reliability of the temporal predictions is further quantified by calculating the credibility maps over time.
An example credibility map is shown in Fig.~\ref{fig:conf_time}(b).
As expected, the BNN is credible across the entire {\it trained} FOV region and less credible over the {\it untrained} region, matching our previous observations.

To quantify the reliability over time, we calculate the averaged credibility over the full FOV, the cell and the background regions [Fig.~\ref{fig:conf_time}(c)].
The  averaged credibility fluctuates within a small range.
The credibility for the cell regions slowly decays over time, which can be explained by that  the temporal dynamics gradually induce more ``dissimilar'' out-of-distribution data. 
Our BNN enables quantifying such ``temporal decorrelation''. 

Next, we zoom in on two small regions where cell divisions undergo over time [Fig.~\ref{fig:conf_time}(d--e)]. 
In both cases, the credibility drops when the cells present significant morphological changes during mitosis, and increases back to the ``normal'' level immediately after the process is over.
More examples are shown in the movie in Visualization 1.
As cells become more globular during mitosis, the phase values grow significantly and often result in phase wrapping errors in the training phase data.
In Fig.~\ref{fig:conf_time}(e), a cell undergoes apoptosis and presents distinct morphological structures.
Similar to our previous observations, the BNN consistently identifies these spatially and temporally rare features by ``flagging'' them as being less credible.

\section{Conclusion}
We have presented a physics-guided DL framework for large-SBP phase imaging.
Our technique enables high-resolution phase inference across a wide FOV using only five asymmetric illumination coded intensity measurements. 
Our results show that this BNN-based technique can effectively learn the underlying physical model.
Once trained, the BNN can robustly solve the phase retrieval problem and is generalizable to different samples. 
Further, we have developed an uncertainty quantification framework that allows critically assessing the reliability of the BNN predictions. 
Specifically, we have applied our UL approach to evaluate the robustness of our illumination coding and DL phase estimation model.
In addition, we have also quantified the effect of common experimental errors using the predicted uncertainties. 
Furthermore, we have showed that applying the UL enables discovering the incompleteness in the training data and quantifying the associated out-of-distribution testing errors.
Finally, the predicted credibility map has shown to be useful in identifying spatially and temporally rare biological phenomena and characterizing the ``temporal decorrelation'' in  dynamic processes. 
We believe this UL framework is widely applicable to many emerging DL-based scientific and biomedical imaging applications where critical assessment to the DL inference is essential.

\section*{Funding Information}
National Science Foundation (1813848),
National Institute of Health (R21GM128020).

\section*{Acknowledgments}
We thank Dr. Ji Yi for providing the samples in our experiments, and Joseph Greene and Alex Matlock for helpful discussions.



\end{document}